# Gate-tunable quantum oscillations in ambipolar $Cd_3As_2$ thin films

Yanwen Liu[1,2], Cheng Zhang[1,2], Xiang Yuan[1,2], Tang Lei[1], Chao Wang[3], Domenico Di Sante[4,5], Awadhesh Narayan[6,7], Liang He[8,9], Silvia Picozzi[4], Stefano Sanvito[6], Renchao Che[3†], Faxian Xiu[1,2†]

[1]State Key Laboratory of Surface Physics and Department of Physics, Fudan University, Shanghai 200433, China

[2]Collaborative Innovation Center of Advanced Microstructures, Fudan University, Shanghai 200433, China

[3]Department of Materials Science and Advanced Materials Laboratory, Fudan University, Shanghai 200433, China

[4]Consiglio Nazionale delle Ricerche (CNR-SPIN), Via Vetoio, L'Aquila, Italy

[5]Department of Physical and Chemical Sciences, University of L'Aquila, Via Vetoio 10, I-67010 L'Aquila, Italy

[6]School of Physics, CRANN and AMBER, Trinity College, Dublin 2, Ireland

[7]Department of Physics, University of Illinois at Urbana-Champaign, Illinois, USA

[8]National Laboratory of Solid State Microstructures, School of Electronic Science and Engineering, Nanjing University, Nanjing 210093, China

[9]Collaborative Innovation Center of Advanced Microstructures, Nanjing University, Nanjing 210093, China

†Correspondence and requests for materials should be addressed to F. X. and R. C. (E-mails: faxian@fudan.edu.cn and rcche@fudan.edu.cn)

**Electrostatic doping in materials can lead to various exciting electronic properties, such as metal-insulator transition and superconductivity, by altering the Fermi level position or introducing exotic phases. $Cd_3As_2$, a three-dimensional (3D) analog of graphene with extraordinary carrier mobility, was predicted to be a 3D Dirac semimetal, a feature confirmed by recent experiments. However, most research so far has been focused on metallic bulk materials that are known to possess ultra-high mobility and giant magnetoresistance but limited carrier transport tunability. Here, we report on the first observation of a gate-induced transition from band conduction to hopping conduction in single-crystalline $Cd_3As_2$ thin films via electrostatic doping by solid electrolyte gating. The extreme charge doping enables the unexpected observation of *p*-type conductivity in a 50 nm-thick $Cd_3As_2$ thin film grown by molecular beam epitaxy. More importantly, the gate-tunable Shubnikov-de Haas (SdH) oscillations and the temperature-dependent resistance reveal a unique band structure and bandgap opening when the dimensionality of $Cd_3As_2$ is reduced. This is also confirmed by our first-principles calculations. The present results offer new insights towards nanoelectronic and optoelectronic applications of Dirac semimetals in general, and provide new routes in the search for the intriguing quantum spin Hall effect in low-dimension Dirac semimetals, an effect that is theoretically predicted but not yet experimentally realized.**

Keywords: topological Dirac semimetal, ionic gating, ambipolar effect, quantum oscillations

**INTRODUCTION**

Dirac materials, such as graphene and topological insulators, have attracted substantial attention owing to their unique band structures and appealing physical properties originated from two-dimensional (2D) Dirac fermions with linear energy dispersion[1–4]. Recently, the existence of three-dimensional (3D) Dirac fermions has been theoretically predicted while several potential candidates including β-$BiO_2$[5], $Na_3Bi$[6] and $Cd_3As_2$[7] were explored as topological Dirac semimetals (TDSs), in which the Dirac nodes are developed via the point contact of conduction-valence bands. By breaking certain symmetries, 3D TDSs could be driven into various novel phases, such as Weyl semimetals[6–8], topological insulators (TIs)[7,8], axion and band insulators[6,8–10], thus providing a versatile platform for detecting unusual states and exploring numerous topological phase transitions.

Among 3D TDSs, $Cd_3As_2$ is considered to be an excellent material due to its chemical stability against oxidation and extremely high mobility[11–14]. Although the electrical, thermal and optical properties of $Cd_3As_2$ have been widely investigated, hampered by the complicated crystal structure its band structure remains a matter of controversy[14,15]. Recently, first-principle calculations have revealed the nature of 3D topological Dirac semimetal state in $Cd_3As_2$[2,7,8]. Soon after the prediction, its inverted band structure with the presence of Dirac fermions was experimentally confirmed[11,13,16–19]. More importantly, beyond the relativistic transport of electrons in bulk $Cd_3As_2$, a theoretically-predicted TI phase may eventually emerge upon the breaking of crystal symetry[7]. Furthermore, thickness-dependent quantum oscillations could be anticipated to arise from arc-like surface states[20]. Such perspective manifests the superiority of $Cd_3As_2$ thin films for the study of the quantum spin Hall effect and the exploration of unconventional surface states in the Dirac semimetals.

Previously, amorphous and crystalline $Cd_3As_2$ films were prepared on various substrates by thermal deposition[21–23], showing SdH oscillations and a quantum size effect[24–26]. However, despite the extensive studies in the past, synthetized $Cd_3As_2$ always exhibits *n*-type conductivity with a high electron concentration, therefore calling for a well-controlled growth scheme and the tunability of carrier density[14,27]. Theory proposed that the chiral anomaly in TDSs can induce nonlocal transport, especially with a large Fermi velocity when the Fermi level, $E_F$, is close to the Dirac nodes[28]. Hence, the ability to modulate the carrier density and $E_F$ in $Cd_3As_2$ plays a vital role for the study of the transport behavior and TDSs-related phase transitions. In view of preserving high mobility in $Cd_3As_2$, the electrostatic doping is an advantageous choice owing to its tunable and defect-free nature compared with the chemical doping.

To modulate a large-area flat film on an insulating substrate, an electric-double-layer transistor (EDLT) configuration was adopted because of its easy device fabrication and high efficiency in tuning the Fermi level, from which a high concentration of carriers can be accumulated on the surface to induce an extremely large electric field[29–34]. In this letter, we demonstrate the tunable transport properties including ambipolar effect and quantum oscillations of wafer-scale $Cd_3As_2$ thin films deposited on mica substrates by molecular beam epitaxy (MBE) (see Method). Our transport measurements reveal a semiconductor-like temperature-dependent resistance in the pristine thin films. Taking advantage of the ionic gating, we are able to tune the Fermi level into the conduction band with a sheet carrier density, $n_s$, up to $10^{13}$ cm$^{-2}$ and witness an evident transition from band conduction to hopping conduction. Moreover, in a certain range of Fermi energy, tunable-SdH oscillations emerge at low temperatures and a transition from electron- to hole-dominated two-carrier transport is achieved by applying negative gate voltage, a strong indication of ambipolar effect, thus demonstrating the great potential of Cd$_3$As$_2$ thin films in electronic and optical applications.

**MATERIALS AND METHODS**

**Sample growth.** Cd$_3$As$_2$ thin films were grown in a Perkin Elmer 425B molecular beam epitaxy system. Cd$_3$As$_2$ bulk material (99.9999%, American Elements Inc.) was directly evaporated onto 2-inch mica substrates by a Knudsen cell. Freshly cleaved mica substrates were annealed at 300 °C for 30 min to remove the molecule absorption. During the growth process, the substrate temperature was kept at 170 °C. The entire growth was *in-situ* monitored by the reflection high-energy electron diffraction (RHEED) system.

**Characterizations of crystal structure of Cd$_3$As$_2$.** The crystal structure was determined by X-ray Diffraction (XRD, Bruker D8 Discovery) and high-resolution Transmission Electron Microscopy (HRTEM, JEOL 2100F, Japan) using a field emission gun. The TEM instrument was operated at 200 KV at room temperature.

**Device fabrication.** The thin films were patterned into standard Hall bar geometry manually. The solid electrolyte was made as follows: LiClO$_4$ (Sigma Aldrich) and poly (ethylene oxide) (PEO, Mw=100,000, Sigma Aldrich) powers were mixed with anhydrous methanol (Alfa Aesar). The solution was stirred overnight at 70 °C and served as the electrolyte. After the application of solid electrolyte, the device was kept at 350 K for 30 min in vacuum to remove the moisture before the transport measurements.

**Device characterizations.** The magneto-transport measurements were performed in a Physical Property Measurement System by Quantum Design with a magnetic field up to 9 T. A home-made measurement system including lock-in amplifiers (Stanford Research 830) and Agilent 2912 source meters was used to acquire experimental data.

**Band structure calculations.** Density functional theory based first-principles calculations were performed for bulk Cd$_3$As$_2$. The resulting bulk Hamiltonian was projected onto a basis of Cd 5*s* and As 4*p* states, using wannierfunctions[35]. The Cd 5*s* orbitals were rigidly shifted by 0.4 eV to match HSE calculations. This ab-initio-derived tight-binding Hamiltonian was then employed to study the system in slab geometries along the [001] direction. Due to the interest here in bulk features, *i.e.*, the evolution of the bulk gap, [001] oriented films were studied for simplicity and qualitative differences for [112] oriented films are not expected. Very recently Cd$_3$As$_2$ has been shown to crystallize into the I41/acd space group (which is a$\sqrt{2} \times \sqrt{2} \times 2$ supercell of the P42/nmc unit cell)[15]. However, the difference in the band structures for the two cells is minimal, and the smaller P42/nmc cell for $Cd_3As_2$ was used to perform the simulations. Density functional theory computations were performed using Vienna Ab-initio Simulation Package (VASP)[36], including spin-orbit coupling. The Perdew-Burke-Ernzerhof parameterization to the exchange-correlation functional was used[37]. A plane wave cutoff of 600 eV was employed, along with a $6 \times 6 \times 3$ Monkhorst-Pack $k$-grid.

**RESULTS AND DISCUSSION**

TEM was carried out to characterize the crystal structure of $Cd_3As_2$. A typical selected-area electron diffraction pattern taken from the same area as the HRTEM image confirms the single crystallinity with the growth face of (112) plane, as shown in Figure 1a and inset. The atom columns cleaving from the original crystal cell mode (Fig. 1e) along (112) plane agree well with that in the HRTEM image (Fig. 1b). The surface morphology of the as-grown thin films was probed by atomic force microscopy (AFM) with a root-mean-square (RMS) of ~ 0.3nm (Fig. 1c). The atomically flat surface is consistent with the 2D growth mode reflected by the streaky RHEED pattern (Figure 1c inset), thus ensuring an ideal solid-liquid interface during the ionic gating process. The top surface can be identified as a series of {112} planes by XRD (Fig.1d), which further confirms the TEM observations.

To carry out low-temperature transport measurements, a ~50 nm-thick $Cd_3As_2$ thin film was patterned into a standard Hall bar configuration with a channel dimension of 2mm × 1mm. A small area of the isolated thin film was left around the channel to serve as a gate electrode. After examining the properties of the pristine sample, a droplet of solid electrolyte was deposited on the device surface to cover the channel area (see Figure 2a). Figure 2b shows the temperature-dependent resistance $R_{xx}$ of the pristine $Cd_3As_2$ thin film prior to the ionic gating process. The negative $dR_{xx}/dT$ suggests semiconducting behavior that is different from the metallic nature of the bulk counterpart[12,19]. The activation energy ($E_a$) is extracted to be 12.45 meV by fitting the Arrhenius plot of $R_{xx}$ at high temperature (from 280 to 350 K) with the equation $R_{xx} \sim exp(E_a/k_BT)$, where $k_B$ is the Boltzmann constant and $T$ is the measurement temperature. The band gap, $E_{gap}$, is roughly estimated to be over 24.9 meV from $E_a$, which is reasonable for the $Cd_3As_2$ thin film of this thickness. The sheet carrier density, $n_s$, at 2K is determined to be $1.5 \times 10^{12}$ $cm^{-2}$ by Hall effect measurements. Such a low carrier density, along with the semiconducting characteristics, indicates that the Fermi level is located inside the bandgap in pristine $Cd_3As_2$ thin films.

With ionic gating, we can efficiently tune the Fermi level in order to achieve two-carrier transport in $Cd_3As_2$. Several as-grown $Cd_3As_2$ thin films have been measured (Supplementary section I). Under positive gate voltage ($0<V_G<0.5$ V, Fig. 2c), $R_{xx}$ shows a negative temperature dependence, indicative of a semiconducting state. Increasing $V_G$ up to 1.2 V, a metallic behavior is witnessed by a change of negative- to positive-temperature dependence. This behavior originates from the fact that the Fermi level has been moved into the conduction band ($V_G \geq 0.5$ V, Fig. 2c). However, when $V_G$ becomes negative, $R_{xx}$ shows a completely negative temperature dependence without metallic behavior owing to the insufficient hole doping (Fig. 2d). Interestingly, the hopping conduction at low temperatures has been observed in this regime, as indicated by the dashed line in Figure 2d. Note that the $R_{xx}$-$T$ curves cross each other at about 50~150 K, suggesting that the Fermi level is closer to the valence band than to the conduction band in this critical temperature range. This gives rise to a hole-dominated transport at low temperatures, which will be investigated in the following section on magneto-transport. The bandgap opening behavior here shows a good agreement with our first-principle calculations. Figure 2e displays the calculated band structure of a typical $Cd_3As_2$ thin film with a thickness of ~ 50 nm. The bulk Dirac cone is fully opened, with a sizable gap larger than 20 meV. This gap falls off with increasing thickness and is very close to zero for a thin film of thickness ~60 nm (see Supplementary Section VIII). This variation in the bulk gap is in reasonable agreement with our experimental results.

In order to further study the gate-tunable $R_{xx}$-$T$ behavior and ascertain the carrier type, magneto-transport measurements were carried out at low temperatures. A clear Hall anomaly at different $V_G$ was observed (see Fig. 3a-d). According to the Kohler's rule[38–40],

$$\frac{R_{xx}(B,T)}{R_{xx}(0,T)} = F\left(\frac{B}{R_{xx}(0,T)}\right) \quad , \tag{1}$$

the magneto-resistance (MR) at different temperatures could be rescaled by the Kohler plot. If there is a single type of charge carrier with the same scattering time at the Fermi surface everywhere, the temperature-dependent Kohler plot of the MR curve would overlap each other.[40] However, there is no field range over which Kohler's rule holds in our experiments (Fig. 3g). Our distinct Kohler curves strongly suggest that two types of carriers with mobilities that have different temperature dependence contribute to the entire transport[40,41]. At high magnetic fields (B≥4T), the slope of Hall resistance $R_{xy}$ approximately equals to $1/[e(n_h-n_e)]$, where $n_h$ and $n_e$ represent the hole and electron density, respectively. Positive $R_{xy}/B$ at high field reveals hole-dominated transport when $V_G$≤-0.9 V (Fig. 3c-d). This Hall slope is sensitive to the Fermi level position and it turns from negative to positive abruptly as $V_G$ changes from -0.6 to -0.9 V, indicating that the Fermi level moves towards the valence band (Fig. 3b-c). On the contrary, at low magnetic fields (B≤2T), the negative $R_{xx}/B$ is attributed to the higher mobility of electrons than that of holes. Upon further decreasing $V_G$ from -0.9 to -2.2 V, the Fermi level moves away from the conduction band and the contribution to $R_{xy}/B$ from electrons at low fields almost vanishes at low temperatures (Fig. 3d, for example, $T$=2 K). This is the result of freezing the residual bulk electrons[42]. Linear $R_{xy}$ with positive slopes suggests a hole-dominated transport in the 50 nm-thick $Cd_3As_2$ thin film.

To quantitatively understand the Hall effect measurements, we employ the two-carrier model with following equation[40,43],

$$R_{xy} = \frac{n_h\mu_h^2 - n_e\mu_e^2 + (\mu_h\mu_e B)^2(n_h - n_e)}{e[(n_h\mu_h + n_e\mu_e)^2 + (\mu_h\mu_e B)^2(n_h - n_e)^2]} \qquad (2)$$

where $n_e$ ($n_h$) and $\mu_e$ ($\mu_h$) represent the carrier density and mobility of electrons (holes), respectively. By preforming the best fit to equation (2), the temperature-dependent mobility and carrier density of both electrons and holes could be acquired. Figure 3e displays the sheet carrier density $n_s$ as a function of gate voltage, where the ambipolar transport characteristic is observed as the holes dominate the negative regime while the electrons prevail in the positive one. The hole density reaches values on the order of $10^{12}$ cm$^{-2}$, comparable to the electron density under positive voltage. Remarkably, the hole mobility rises from ~500 to ~800 cm$^2$V$^{-1}$s$^{-1}$ as the gate changes from -0.8 to -2.2 V, which is consistent with the transition from two-carrier to hole-dominant transport. In contrast, the electron mobility reaches ~3000 cm$^2$V$^{-1}$s$^{-1}$ when the Fermi level locates in the conduction band (Supplementary Fig. S3). Presumably, the hole carriers with low band velocity could suffer severe impurity scattering as observed in scanning tunneling microscopy experiments[18]. So owing to the low mobility, it is difficult to observe the SdH oscillations from the hole carriers. According to the equation $\sigma=ne\mu$, the ratio of conductivity $\sigma_p/\sigma_n$ can be calculated for each gate voltage, and in general it decreases as the temperature increases (Fig. 3f), suggesting the increasing component of electron conduction in the channel. The ratio crosses 1 at about 60~100 K (dashed lines in Fig. 3f), which is reasonably consistent with the previous $R_{xx}$-$T$ analysis (Fig. 2d). Moreover, the ratio of conductivity $\sigma_p/\sigma_n$ exceeds 9 at 2 K for the gate voltage of -2.2 V, demonstrating the hole-dominant transport here. A detailed discussion of two-carrier transport is presented in Supplementary Section IV.

Quantum oscillation serves as an effective way to probe the Fermi surface of band structure[43,44]. Under positive $V_G$, the SdH oscillations can be well-resolved as the Fermi level enters the conduction band, leading to the increase of electrons which adopt a relatively high mobility. Figure 4a shows gate-dependent SdH oscillations of Cd$_3$As$_2$ at 4 K. According to the linear and negative slope of $R_{xy}$/$B$ (Fig. 3a), electrons are predominant in the transport leading to the SdH oscillations at high magnetic fields. To fundamentally understand the SdH oscillations at different $V_G$, we calculate the oscillation frequency $F$ by taking the periodic maxima and minima of $R_{xx}$. From the equation $F= (\phi_0/2\pi^2)A_F$, where $\phi_0=h/2e$, we can obtain the cross-section area of the Fermi surface (FS) $A_F$. As $V_G$ changes from 0 to 1.2 V, $F$ increases from 18.1 to 42.5 T, translating to the variation of $A_F$ from $1.72\times10^{-3}$ to $4.05\times10^{-3}$ Å$^{-2}$. The enlargement of FS suggests that the Fermi level moves deeper into the conduction band as $V_g$ becomes larger. According to $A_F=2\pi k_F^2$, the Fermi vector of $k_F$ can be extracted as summarized in Table I. In contrast, owing to the low mobility of holes, SdH oscillations were not detected under negative gate voltage when the Fermi level is near the valence band.

The SdH amplitude as a function of temperature can be analyzed to obtain more important parameters of the carrier transport. Here we particularly focus on the SdH oscillations under $V_g$=0 V. The temperature-dependent amplitude $\Delta R_{xx}$ (Fig. 4b) is described by $\Delta R_{xx}(T)/R_{xx}(0)=\lambda(T)/\sinh(\lambda(T))$, and the thermal factor is given by $\lambda(T)=2\pi^2k_BTm_{cyc}/(\hbar eB)$, where $k_B$ is the Boltzmann's constant, $\hbar$ is the reduced plank constant and $m_{cyc}=E_F/v_F^2$ is the cyclotron mass. By performing the best fit to the $\Delta R_{xx}(T)/\Delta R_{xx}(0)$ equation, $m_{cyc}$ is extracted to be 0.029 $m_e$. Using the equation $v_F=\hbar k_F/m_{cyc}$, we can obtain the Fermi velocity $v_F=9.27\times10^5$ m/s and the Fermi energy $E_F$=143 meV. From the Dingle plot, the transport life time, $\tau$, the mean free path $l=v_F\tau$, and the cyclotron mobility $\mu_{SdH}=e\tau/m_{cyc}$ could be estimated to be $1.25\times10^{-13}$ s, 116 nm and 7537 cm$^2$V$^{-1}$s$^{-1}$, respectively. By performing the same analysis for other gate voltages, we can extract all the physical parameters (Fig. 4e), as provided in Table I.

As the gate voltage changes from 0 to 1.2 V, the Fermi energy increases from 143

to 254 meV after applied solid electrolyte, showing the lifting of the Fermi level into the conduction band (Table I). Also the lifetime and Fermi velocity give remarkable values approaching ~$10^{-13}$ s and $10^6$ cm/s, respectively, which are approximate to previous transport results of the bulk material[12,19]. With continuous electron doping by applying even larger positive $V_g$, the Fermi level goes further into the conduction band and the amplitude of the SdH oscillations gets significantly weakened and finally vanishes, suggesting increasing scattering deep into the conduction band (Fig. 4a and d; Supplementary Fig. S10). To further understand the gate-tunable SdH oscillations, Berry's phase has been evaluated from the Landau fan diagram as shown in Fig. 4c. Here, we assign integer indices to the $\Delta R_{xx}$ peak positions in 1/B and half integer indices to the $\Delta R_{xx}$ valley positions[44]. According to the Lifshitz-Onsager quantization rule[44] : $A_F \frac{\hbar}{eB} = 2\pi \left( n + \frac{1}{2} - \frac{\Phi_B}{2\pi} \right) = 2\pi(n + \gamma)$, the Berry's phase $\Phi_B$ can be extracted from the intercept, γ, in the Landau fan diagram by $\gamma = \frac{1}{2} - \frac{\Phi_B}{2\pi}$. For nontrivial π Berry's phase, γ should be 0 or 1, as shown in previous experiments for bulk $Cd_3As_2$[19]. In our samples, under different gate voltages the intercept remains close to 0.5, indicating a trivial zero Berry's phase. The presence of zero Berry's phase reveals that the SdH oscillations mainly derive from the high mobility bulk conduction band. With the dimensionality reduced from bulk to thin film, $Cd_3As_2$ exhibits a transition from topological Dirac semimetal to trivial band insulator[7]. The Dirac point vanishes following the band gap opening. Even so, the advantage of high mobility in the $Cd_3As_2$ bulk material is preserved along with the small effective mass and long lifetime although a large linear MR[12] is absent here. The detailed magneto-transport mechanism for both the bulk and thin film of $Cd_3As_2$ remains elusive at this stage and it deserves further investigation.

Angular dependent measurements were also employed for each gate voltage showing SdH oscillations. As the magnetic field is tilted away from the sample normal, the amplitude of the SdH oscillations starts to decrease as long as the angle passes 45° (Supplementary Section VI), presumably attributed to the anisotropic Fermi surface arising from the quantum confinement in the normal direction[7]. This may explain the deviation from the bulk materials in which the SdH oscillations were observable from 0° to 90°[12]. Furthermore, we use polar plots to identify the anisotropy of the MR[12]. Below 1 T, the MR is nearly isotropic under different gate voltage (Fig. 5a). As the magnetic field increases, the polar plots assume a dipolar pattern (Fig. 5b). When increasing further the gate voltage, the dipolar component decreases, giving the trend of crossover to isotropic behavior (Fig. 5c). We note that with increasing the carrier density, it needs larger magnetic field to make the Fermi surface occupy the same Landau level. Indeed, the polar plot of 1.2 V at 9 T displays a similar pattern to that of 0 V at 5 T (Fig. 5b-c), indicating the reduction of anisotropy by either lifting up $E_F$ or decreasing $B$. (Fig. 5c). Inspired by the previous transport analysis, when the Fermi level moves into the conduction band, the anisotropy could be reduced with the enhancement of the scattering processes as evidenced by the decrease of both Hall and quantum mobility. The former one is affected by large angle scattering, *i.e.,* the transport scattering, while the latter is influenced by both small and large angle scattering (Supplementary Fig. S3 and Table I). According to the study of bulk materials[12], the anisotropy mainly originates from the anisotropic transport scattering. With increasing gate voltage, the quantum mobility decreases from ~8000 to ~2700 cm$^2$V$^{-1}$s$^{-1}$ while the Hall mobility decrease from ~3600 to 2500 cm$^2$V$^{-1}$s$^{-1}$. The more rapid reduction of the quantum lifetime reduces the role of transport scattering, leading to the reduction of the anisotropy. This behavior can also be verified by the Kohler's plots (Supplementary Section IV).

**CONCLUSION**

In conclusion, taking advantage of the high capacitance of the solid electrolyte, we demonstrate for the first time a gate-tunable transition of band conduction to hopping conduction in single-crystalline Cd$_3$As$_2$ thin films grown by MBE. The two-carrier transport along with the controllable $R_{xx}$-$T$ suggests that Cd$_3$As$_2$ can generate a small band gap as the system reduces dimensionality. Importantly, SdH oscillations emerge when the Fermi level enters into the conduction band with high electron mobility. Thus, Cd$_3$As$_2$ thin film systems hold promise for realizing ambipolar field effect transistors and for observing intriguing quantum spin Hall effect.

**CONFLICT OF INTEREST**

The authors declare no conflict of interest.

ACKNOWLEDGEMENTS

This work was supported by the National Young 1000 Talent Plan, Pujiang Talent Plan in Shanghai, Ministry of Science and Technology of China (973 Project Nos.2013CB923901) and National Natural Science Foundation of China (61322407, 11474058, 61474061, 11274066, U1330118). Part of the sample fabrication was performed at Fudan Nano-fabrication Laboratory. We thank Mr. Yijun Yu and Professor Yuanbo Zhang for great assistance on solid electrolyte. We acknowledge Professor Shiyan Li for the inspiring discussions. A.N. acknowledges support from the Irish Research Council under the EMBARK initiative. S.S. acknowledges support from the European Research Council (QUEST project).

**FIGURE LEGENDS**

**Figure 1 | Characterizations of as-grown Cd$_3$As$_2$ thin films. a,** A typical HRTEM image of Cd$_3$As$_2$ thin films, revealing a single crystalline structure. Inset: selected-area electron diffraction pattern. **b,** The amplified image in the red box in **a**, perfectly agreeing with the atom columns cleaving from the original crystal cell mode of Cd$_3$As$_2$ in **e** along (112) plane. **c,** AFM image of the thin film surface. The RMS was determined to be ~0.3 nm. Inset: *in-situ* RHEED pattern during growth. **d,** XRD spectrum. The marked peaks are the typical XRD patterns from Cd$_3$As$_2$ with a (112) plane of sample surface, while other peaks come from the mica substrate. **e,** Simulated crystal structure of Cd$_3$As$_2$.

**Figure 2 | Electric transport of ~50 nm-thick Cd$_3$As$_2$ thin film with solid electrolyte gating. a,** A schematic view of solid electrolyte gated Cd$_3$As$_2$ device structure. The inset shows the geometrical configuration of the magnetic field. **b,** Temperature-dependent longitude resistance $R_{xx}$ of Cd$_3$As$_2$ before the application of the solid electrolyte. Inset: the Arrhenius plot of $R_{xx}$. **c,** Temperature-dependent $R_{xx}$ with different positive gate voltages $V_g$, displaying a gate-induced metallic behaviour. **d,** Temperature-dependent $R_{xx}$ of gated Cd$_3$As$_2$ Hall bar device with different negative gate voltage $V_g$. The gate-induced hole accumulation in *n*-type Cd$_3$As$_2$ results in semiconducting-like $R_{xx}$-$T$ behaviour. The dashed curve here represents the fitting of the hopping conduction at low temperatures. **e,** The electronic band structure of the Cd$_3$As$_2$ thin film with a thickness of ~50 nm.

**Figure 3 | Temperature- and gate-dependent Hall resistance $R_{xy}$ of ~50 nm-thick Cd₃As₂ thin film. a,** $R_{xy}$ under -0.5 V (gate voltage), indicative of electron-dominated *n*-type conductivity. **b,** $R_{xy}$ under -0.6 V, showing a nonlinear behaviour originated from two-carrier transport owing to the gate-induced holes. **c,** $R_{xy}$ under -0.9 V. The Cd₃As₂ channel undergoes a transition from electron- to hole- dominated transport as evidenced by the change of slope at B≥3T. **d,** $R_{xy}$ under -2.2 V. The holes are dominant in Hall resistance. **e,** Gate-dependent sheet carrier density. It implies the ambipolar transport. The hole carrier density was extracted from the fits to the two-carrier transport model. Electron carrier density was obtained from the Hall effect measurements. The graduated background represents the amount and type of carriers, blue for holes and red for electrons. **f,** Temperature-dependent conductance ratio $\sigma_n/\sigma_p$. The dashed line marks $\sigma_n/\sigma_p=1$. **g,** The Kohler's plots of the MR curves at the gate voltage of -0.9 V. The non-overlapping behavior with the non-linear Hall data suggests unambiguously two-carrier transport.

**Figure 4 | SdH oscillations of Cd₃As₂ thin films. a,** Gate-dependent SdH oscillations at 4 K. The amplitude decreases as the gate voltage increases. Also the critical magnetic field that the oscillations start shifts towards higher field with higher gate voltage. **b,** Temperature-dependent SdH oscillations at 0 V. **c,** Landau level index n with respect to 1/B under different gate voltages. Integer indices denote the $\Delta R_{xx}$ peak positions in 1/B and half integer indices represent the $\Delta R_{xx}$ valley positions. The intercepts are close to 0.5. **d,** Temperature-dependent amplitude of SdH oscillations under different gate

voltages. With the best fit, the effective mass was obtained. **e,** Effective mass and quantum lifetime as a function of carrier density obtained from the Hall effect measurements. The effective mass slightly increases with increasing carrier density while the quantum lifetime shows the opposite trend.

**Figure 5| Polar plots of the angle variation of MR. a,** Polar plots of the low field MR for selected field under different gate voltages. The MR becomes nearly isotropic below 1T. **b,** Polar plots of the high field MR for selected field under different gate voltages. The MR pattern changes from an isotropic to a dipolar form with increasing magnetic field, suggesting the anisotropy of MR under high field. **c,** Polar plots of the MR fixed at selected field under different gate voltages. Under larger gate voltage, the $E_F$ becomes large and the anisotropy is reduced.

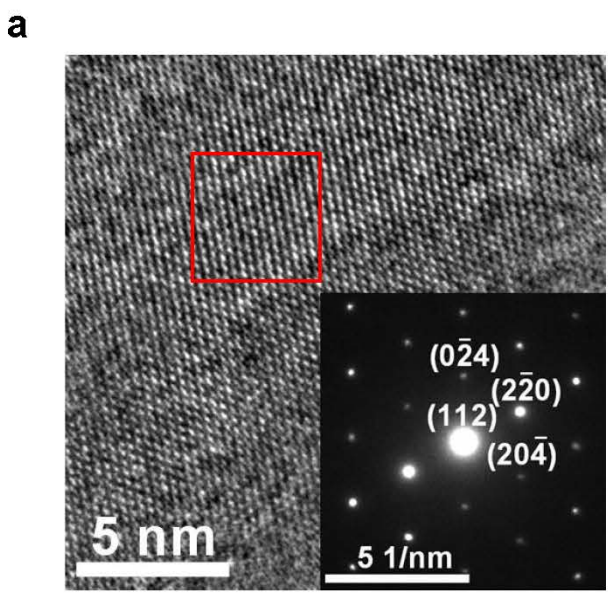
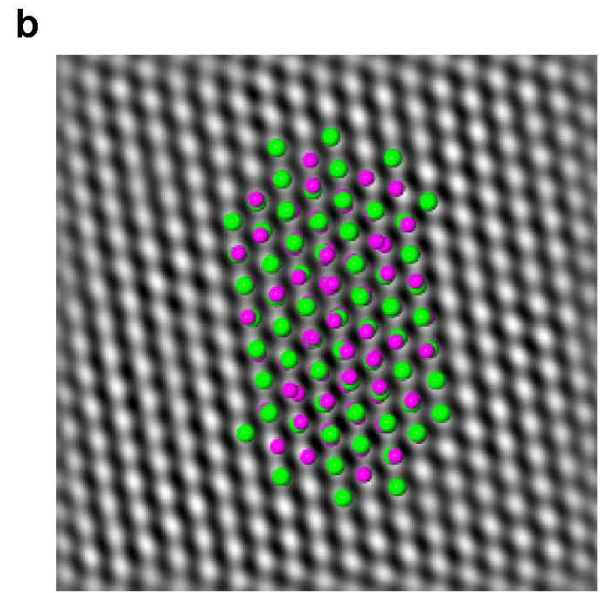
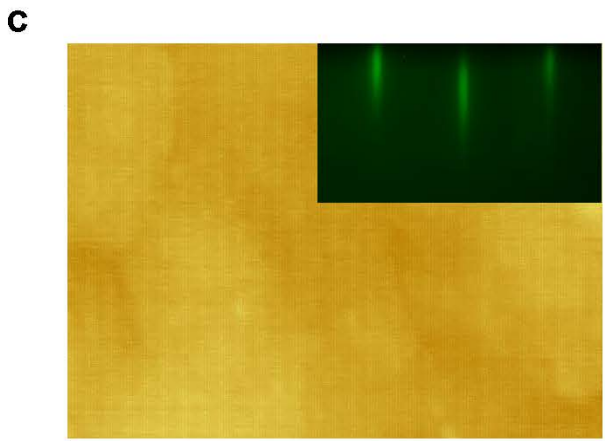
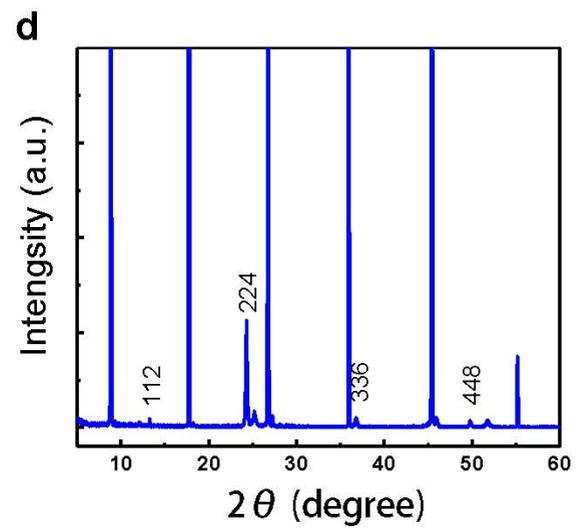
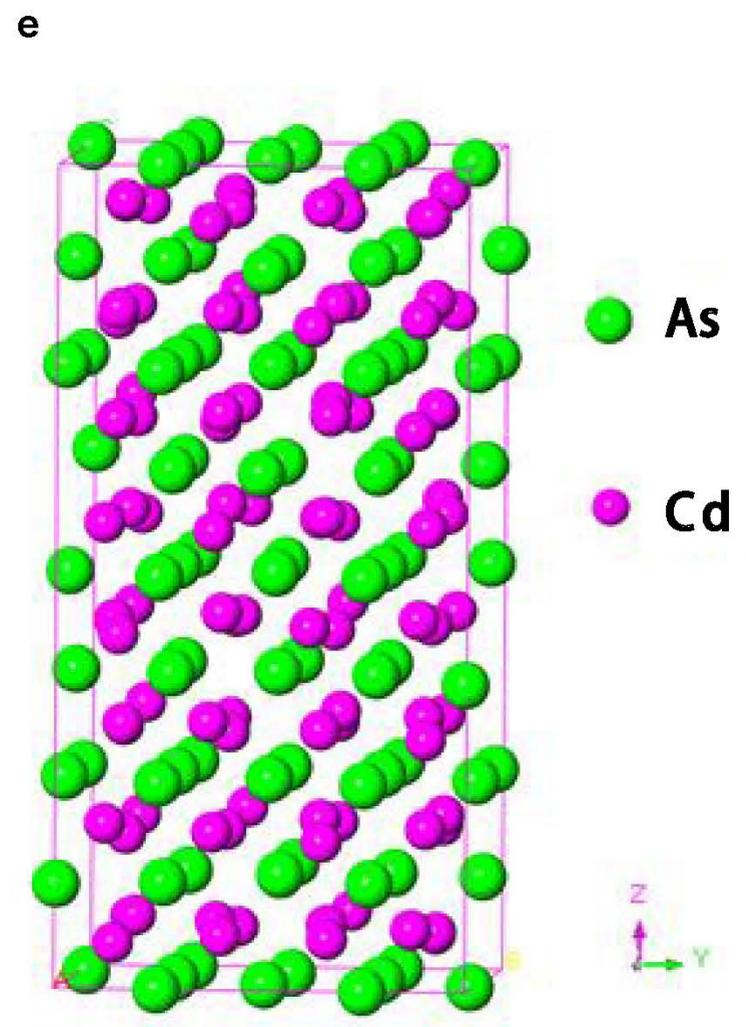

Yanwen Liu *et al*. Figure 1

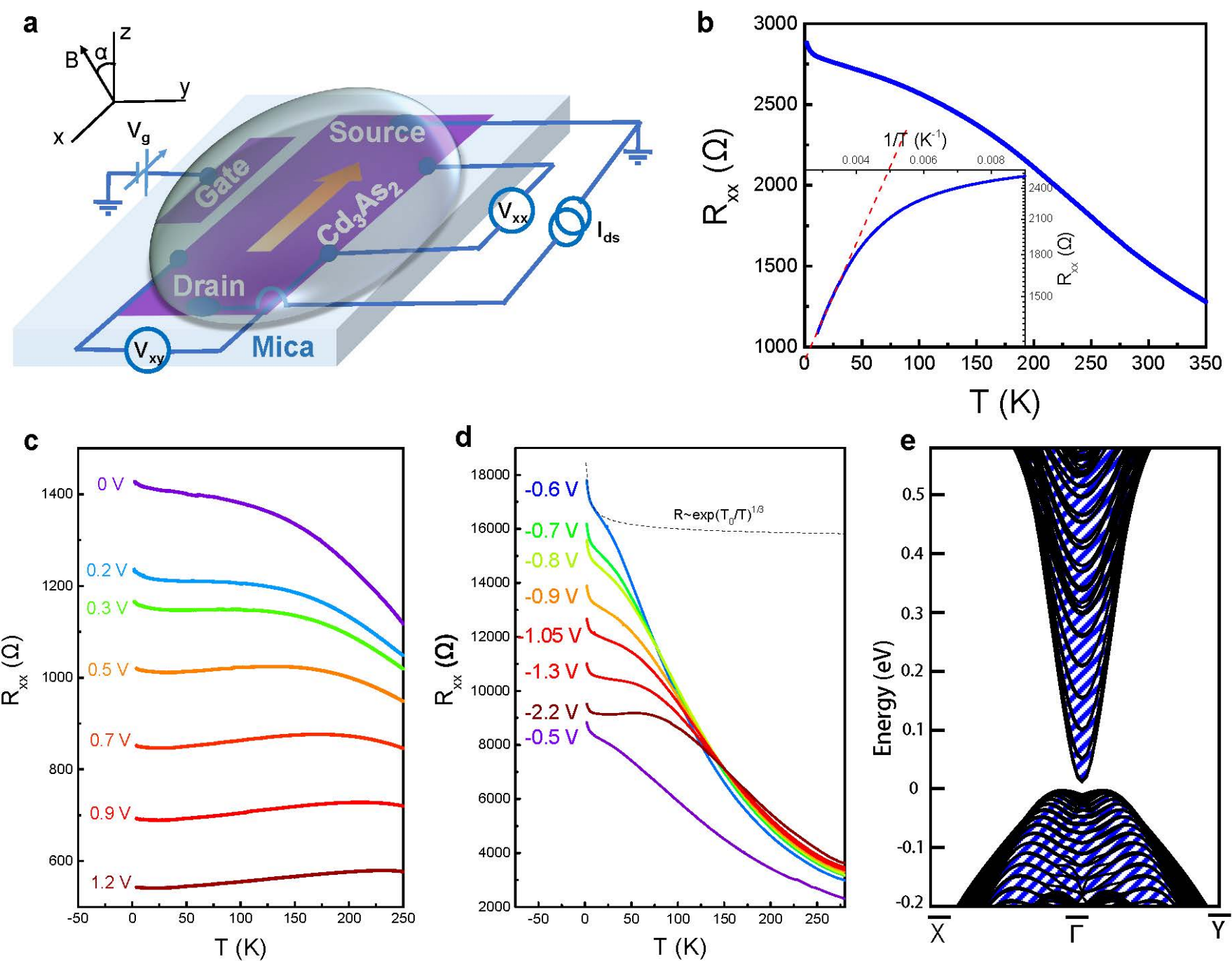

Yanwen Liu *et al.* Figure 2

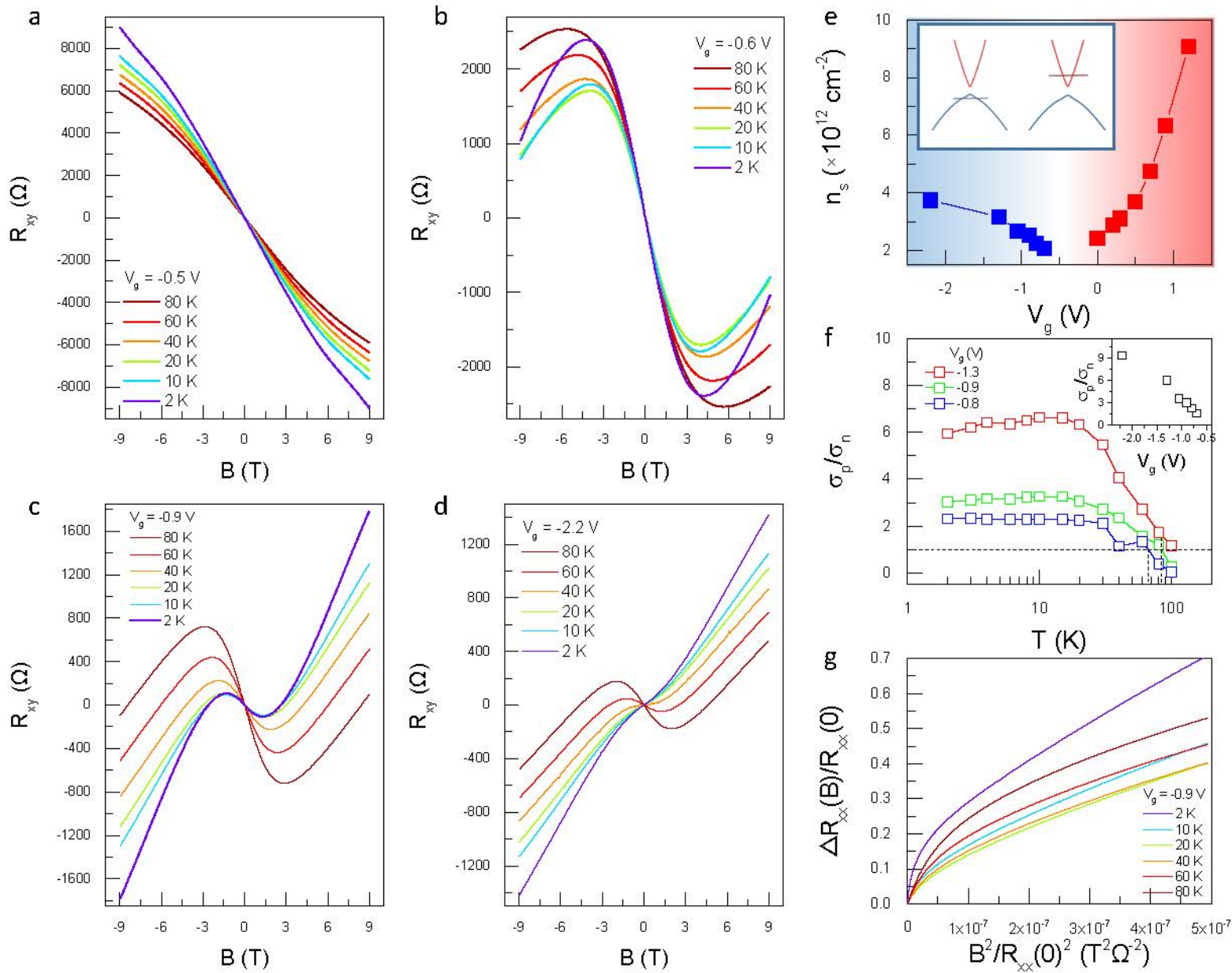

Yanwen Liu *et al.* Figure 3

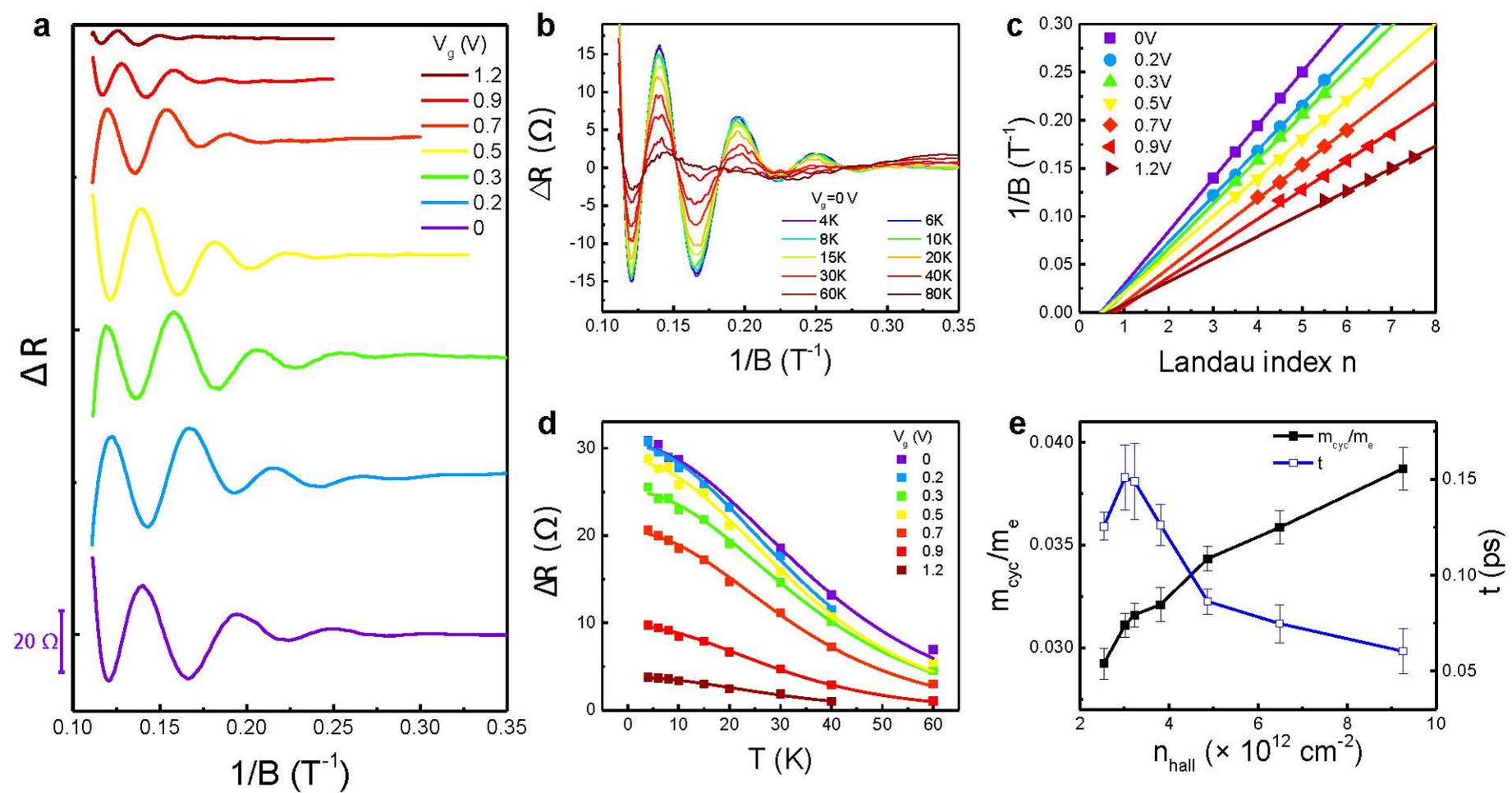

Yanwen Liu *et al.* Figure 4

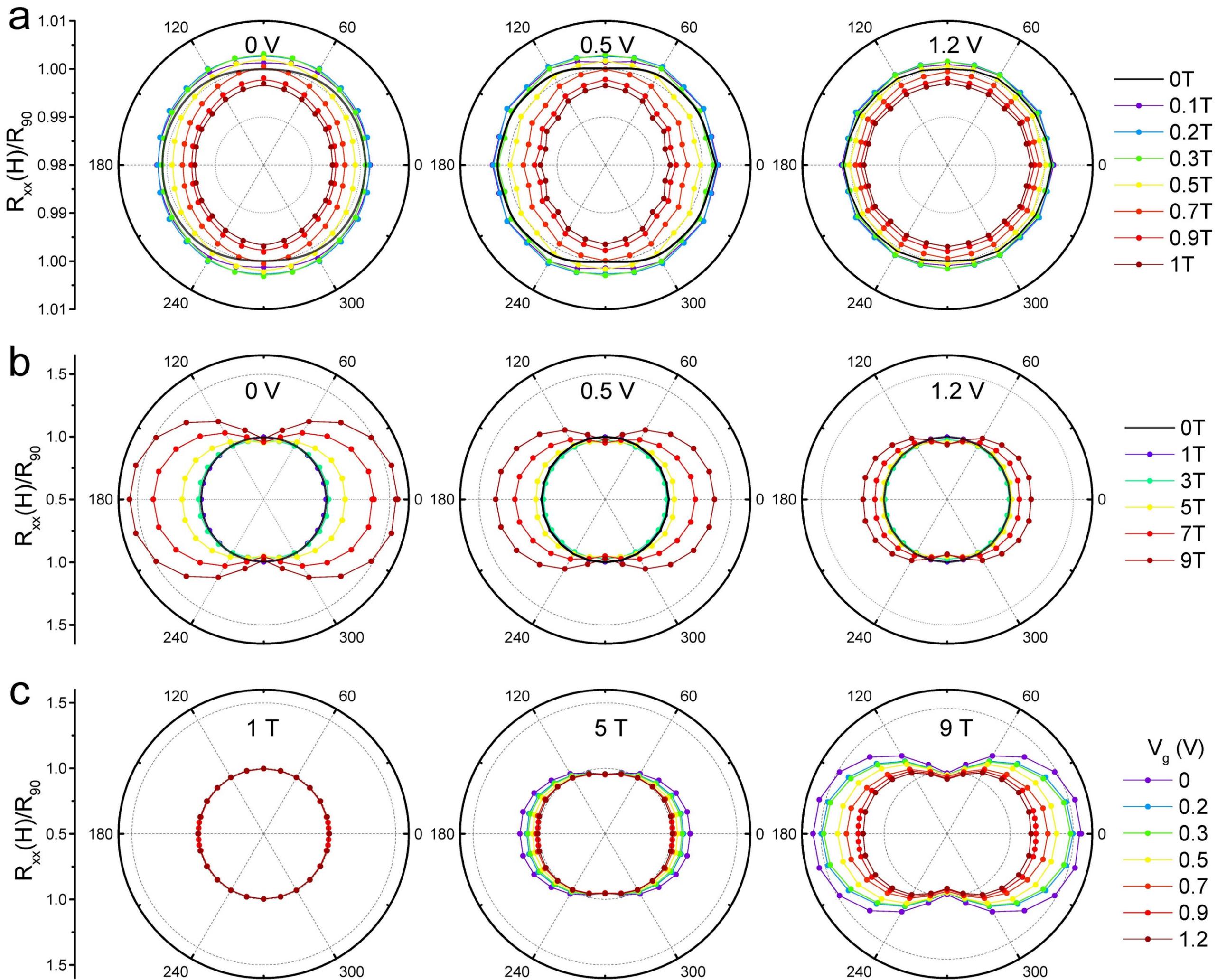

Yanwen Liu *et al.* Figure 5

### Table I | Estimated parameters from the SdH oscillations at T=4 K

| $V_g$(V) | $F_{SdH}$(T) | $k_f$(Å) | $m_{cyc}(m_e)$ | $v_F(10^5$ m/s) | $E_f$(meV) | $t(10^{-13}$ s) | $l$(nm) | $\mu_{SdH}$(cm$^2$V$^{-1}$s$^{-1}$) |
|---|---|---|---|---|---|---|---|---|
| 0   | 18.1 | 0.0234 | 0.0292 | 9.3 | 143 | 1.25 | 116 | 7537 |
| 0.2 | 20.6 | 0.0252 | 0.0311 | 9.4 | 155 | 1.51 | 141 | 8524 |
| 0.3 | 21.2 | 0.0254 | 0.0317 | 9.3 | 157 | 1.49 | 138 | 8241 |
| 0.5 | 24.9 | 0.0275 | 0.0321 | 9.9 | 180 | 1.26 | 125 | 6910 |
| 0.7 | 27.7 | 0.0290 | 0.0343 | 9.8 | 187 | 0.86 | 84  | 4416 |
| 0.9 | 32.9 | 0.0316 | 0.0358 | 1.0 | 213 | 0.75 | 76  | 3664 |
| 1.2 | 42.5 | 0.0359 | 0.0387 | 1.1 | 254 | 0.60 | 65  | 2736 |

Yanwen Liu *et al.* Table 1